\documentclass[12pt,preprint]{aastex}

\shorttitle{}
\shortauthors{Pereyra et al.}

\begin{document}


\title{The deceleration of nebular shells in evolved planetary nebulae\footnote{The observations reported herein were acquired at the Observatorio Astron\'omico Nacional in the Sierra San Pedro M\'artir (OAN-SPM), B. C., Mexico.}}


\author{Margarita Pereyra, Michael G. Richer \& Jos\'e Alberto L\'opez}
\affil{OAN, Instituto de Astronom\'\i a, Universidad Nacional Aut\'onoma de M\'exico, \\ Apartado Postal 106, 22800 Ensenada, BC, M\'exico}
\email{mally@astrosen.unam.mx, richer@astrosen.unam.mx \& jal@astrosen.unam.mx}



\begin{abstract}

We have selected a group of 100 evolved planetary nebulae (PNe) and study their kinematics based upon spatially-resolved, long-slit, echelle spectroscopy. The data have been drawn from the San Pedro 
M\'{a}rtir Kinematic Catalogue of PNe \citep{Lopez12}. The aim is to characterize in detail the global kinematics of PNe at advanced stages of evolution with the largest sample of homogenous data used to date for this purpose. The results reveal two groups that share kinematics, morphology, and photo-ionization characteristics of the nebular shell and central star luminosities at the different late stages under study.  The typical flow velocities we measure are usually larger than seen in earlier evolutionary stages, with the largest velocities occurring in objects with very weak or absent [\ion{N}{2}] $\lambda$6584 line emission, by all indications the least evolved objects in our sample.  The most evolved objects expand more slowly.  This apparent deceleration during the final stage of PNe evolution is predicted by hydrodynamical models, but other explanations are also possible. These results provide a template for comparison with the predictions of theoretical models.
\end{abstract}





\section{Introduction}\label{sec_introduction}

Over the last three decades, many efforts have focused on describing the kinematic behavior of the nebular shell throughout the life of a planetary nebula (PN), from the interacting stellar wind (ISW) model proposed by \cite{Kwok82} to the recent hydrodynamical models \citep{Villaver02, Perinotto04, Schonberner05a, Schonberner05b}.  According to these models, the kinematic evolution of nebular shells in PNe depends basically on 2 factors: the mass loss on the AGB and the energy provided by the central star (CS) through its wind and radiation field.  
Initially, the nebular shell expands with the velocity of the wind ejected during the AGB phase ($\sim$ 10 km s$^{-1}$). As the central star evolves and contracts, increasing its temperature and surface gravity, its wind velocity also increases.  The now fast wind of the CS expands against the previously ejected slow wind from the AGB stage in a momentum-conserving phase, forming a high density region between them. 
As the CS evolves to higher temperatures, its UV radiation becomes sufficient to ionize the surrounding material and we see the nebular shell in optical line emission. The CS's radiation field drives an ionization front through the nebular shell, increasing the observed outflow velocity of the bulk of the mass \citep[by $\sim$ 5 km s$^{-1}$;][]{Chevalier97, Perinotto04}.
Eventually, the energy accumulated behind the shock between the two winds is so high that radiation cooling becomes inefficient and an energy-driven hot bubble forms inside the shell. The pressure from the hot bubble imposes an additional acceleration upon the bulk of the nebular shell, of the order of 15 km s$^{-1}$ \citep{Kwok00}, depending upon the metallicity and probably AGB wind properties \citep[Fig. 13,][]{Schonberner10}.  What happens at the latest stages of evolution is less clear but theoretical work suggests that, after a quick drop in both CS luminosity and stellar wind energy, the hot bubble cools allowing a deceleration of the nebular shell and a possible backflow of the inner parts of the shell towards the CS \citep{GS06}. 

Our current theoretical understanding of the kinematics of the nebular shells in PNe clearly predicts an evolution of the kinematics with time \citep{Kwok82, Mellema94, Villaver02, Perinotto04, Schonberner05a, Schonberner05b}.  \citet{Richer08, Richer10} clearly recover the early evolution.  Some studies provide observational support for the predictions of theoretical models during the latest stages of PN evolution \citep{B&S74,Chu84,H&W90,Kaler90,Phillips02,Pena03}. However, their results are often based on small samples where the relation between the evolution of the CS  and the kinematic evolution of the nebular shell is difficult to assess.


In order to gain a better understanding of the evolution of the nebular shell for PNe in advanced stages of evolution, we analyze the kinematics of a large sample (100 PNe) of evolved PN drawn from the San Pedro Martir Kinematic Catalogue of Planetary Nebulae \citep{Lopez12}, the largest homogeneous database of its kind existing today. The sample's size and homogeneity provide a greater opportunity to tie the theory to observations from a solid statistical basis. The selection criteria and the resulting data are described in \S \ref{sec_observations}. In \S \ref{sec_sample_results}, we present the statistical results for the kinematic behavior of the entire sample and the analysis of two distinct evolutionary stages we find.  In this section, we also present the analysis for a subsample of objects for which we have adopted stellar parameters (L/L$_{\odot}$, T$_{eff}$, and distance) from \cite{Frew08}. Finally, in \S \ref{sec_discussion} 
and \S \ref{sec_conclusions} we discuss the results and summarize our conclusions.

\section{The Evolved PN Sample}\label{sec_observations}

\subsection{Observations and Selection criteria}\label{sec_sample_def}

Every object in this sample was selected from the San Pedro Martir Kinematic Catalogue of Planetary Nebulae \cite[SPM catalogue;][]{Lopez12}. This catalogue contains high resolution spectra of 614 galactic and 211 extragalactic PNe, observed over several years from the Observatorio Astron\'omico Nacional in the Sierra de San Pedro M\'artir (OAN-SPM) in Ensenada, B.C., M\'exico. The observations were obtained with the Manchester Echelle Spectrograph, a long slit, echelle spectrograph without cross-dispersion \citep{meaburnetal1984,meaburnetal2003}. Narrow-band filters isolate orders 87 and 114 containing H$\alpha$ (including [\ion{N}{2}] $\lambda$$\lambda$6548,6584) and [\ion{O}{3}] $\lambda$5007, respectively. 
The spectral resolution is 11 km s$^{-1}$, easily resolving the internal kinematics of typical PNe. The spectra are calibrated using exposures of a ThAr lamp, with an accuracy of $\pm1$ km s$^{-1}$ when converted to radial velocity. 

\begin{figure}[]
\begin{center}
\includegraphics [width=6.0in]{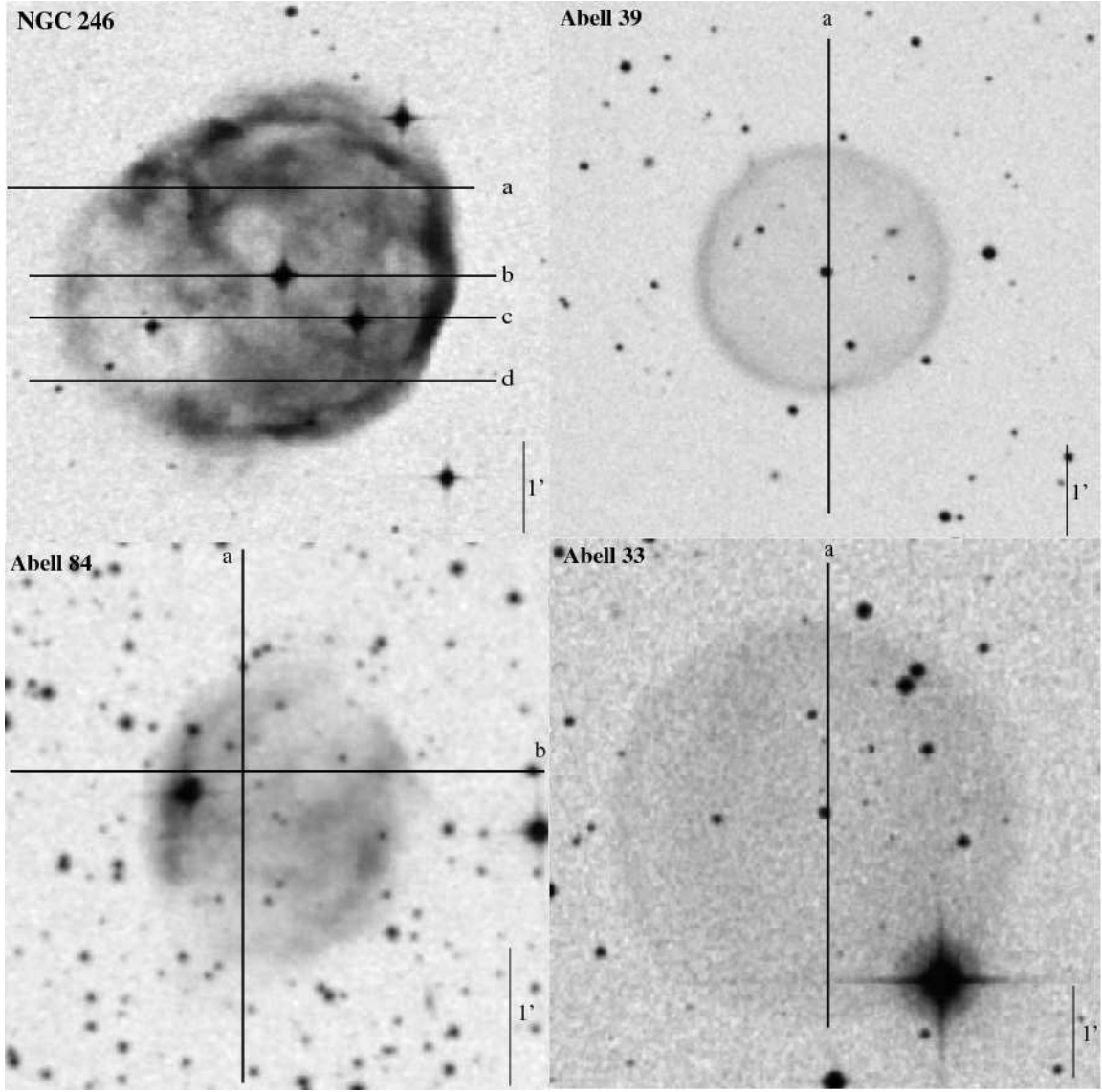} 
\caption{We present images, for NGC 246, Abell 39, Abell 84, and Abell 33 to illustrate our morphological groups.  NGC 246 and Abell 39 are classified as Mature PNe, because they show some remnant structure, possibly reflecting an ionization front.  On the other hand, Abell 84 and Abell 33 are classified as HE PNe due to their completely diffuse and structureless morphology. The solid lines in each image indicate the slit position used for each object.}
   \label{figure01}
\end{center}
\end{figure}

We selected a sample of evolved PNe that are extended and spatially resolved  from the SPM catalogue classifying as an evolved PN those objects with (i) essentially roundish morphology, without complex inner structures  like FLIERS, jets, point symmetry, bipolarity, knots, extended halos or multiple shells, (ii) low surface brightness, and (iii) mostly no bright rims or edges.  These objects should be those that may be compared most easily with existing 1-D hydrodynamical models.  Within the selected group we further subdivided them into moderately evolved or mature (M) and highly evolved (HE) nebulae. The highly evolved nebulae display very low surface brightness, no inner structure and no bright rim whereas the mature nebulae may show some structure and/or some bright edges but apart from that have the characteristics mentioned above of well developed PNe.
This, somewhat arbitrary classification, is complemented with data for the central stars, when available, as explained below.
Representative examples of both classes of evolved PNe from our sample are showed in Fig. \ref{figure01}.

The complete sample of 100 objects is presented in Table \ref{table01}. The first two columns show the common and PNG name for each object.  Columns 3 through 5 present our measurements of the bulk flow velocity (V$_{flow}$), from [\ion{N}{2}] $\lambda$6584, H$\alpha$, and [\ion{O}{3}] $\lambda$5007. Columns 6 and 7 indicate when the [\ion{N}{2}] $\lambda$6584 emission is present and the evolutionary group, respectively.  Columns 8, 9 and 10 list the temperature and luminosity for the central star \citep{Frew08} and the \ion{He}{2} $\lambda$4686 intensity \citep{Tylenda94}, respectively. 

\subsection{Measurements}\label{subsec_nebular_par}

\begin{figure*}[]
\begin{center}
\includegraphics [width=3.0in]{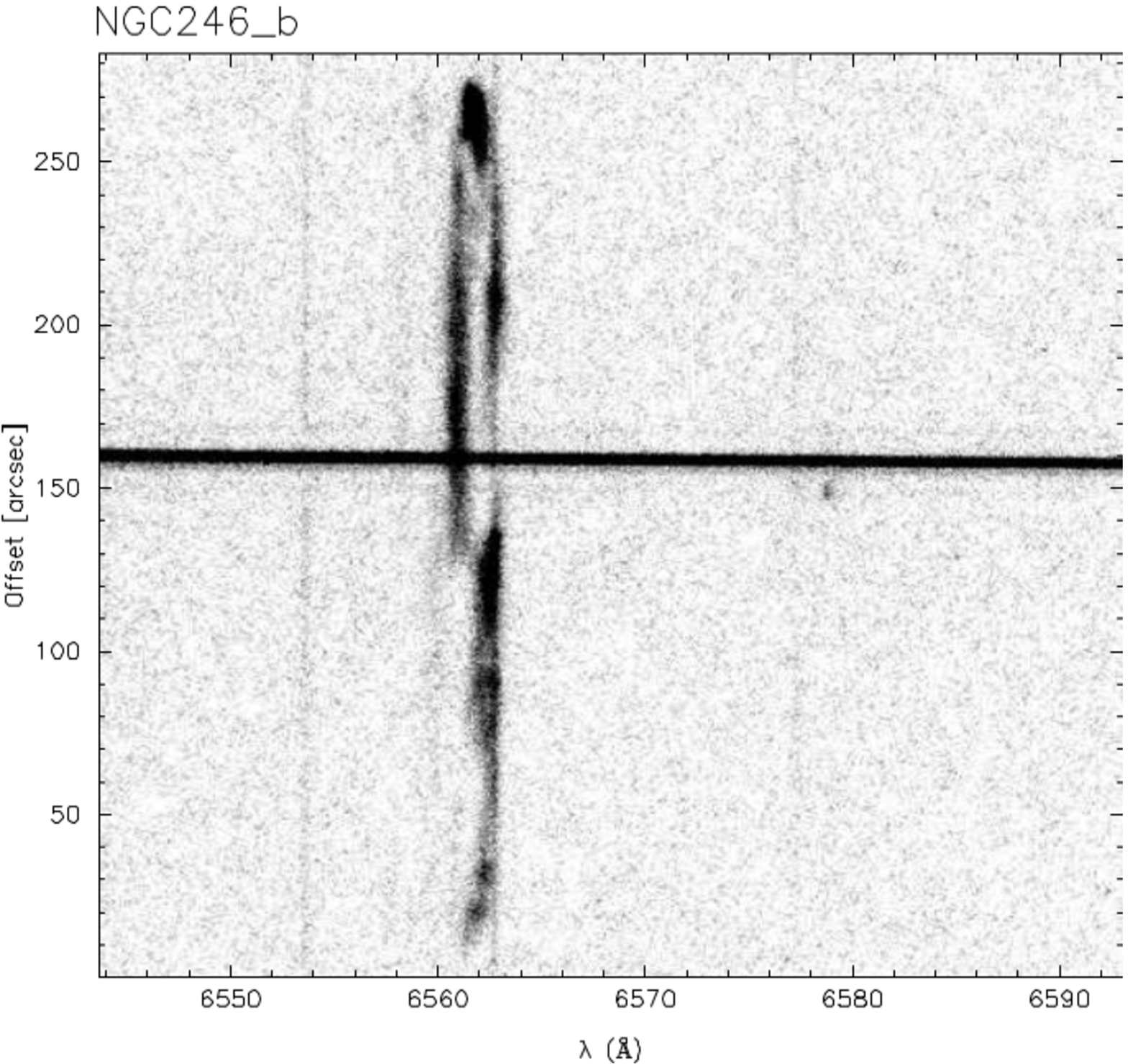} 
\includegraphics [width=3.4in]{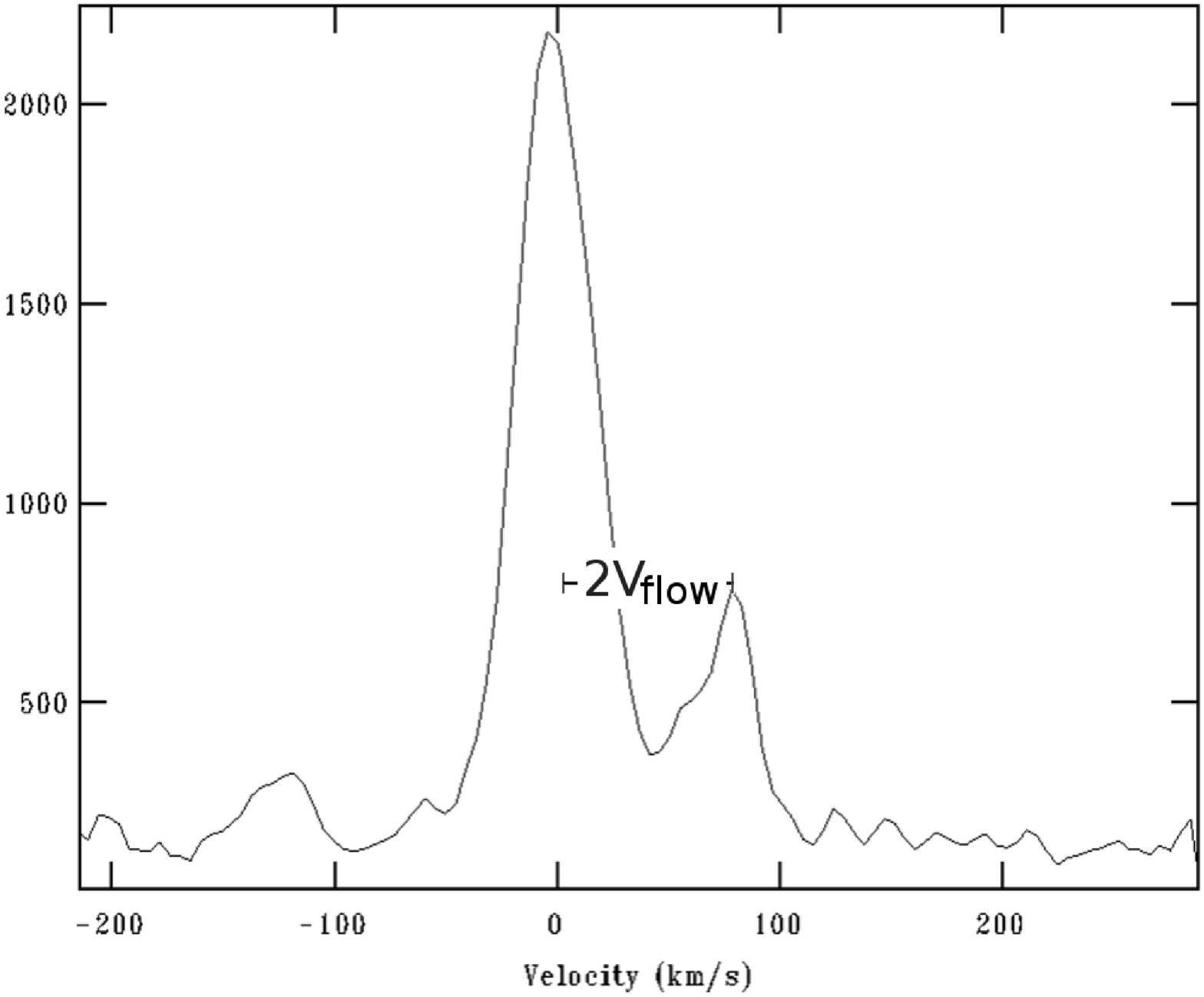} 
\includegraphics [width=3.0in]{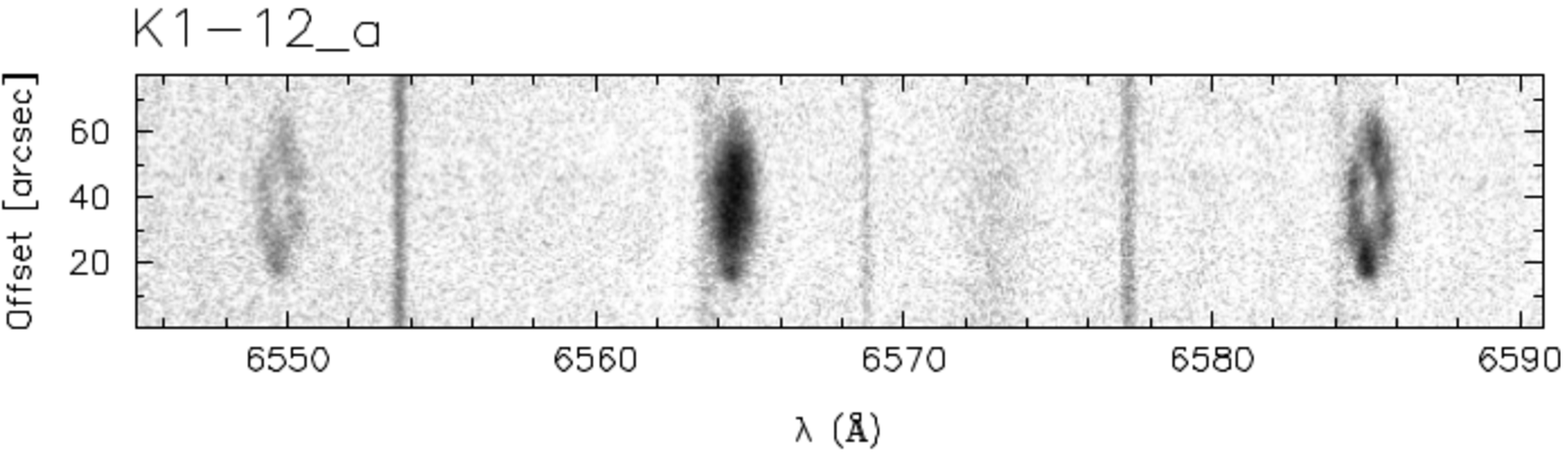}
\includegraphics [width=3.4in]{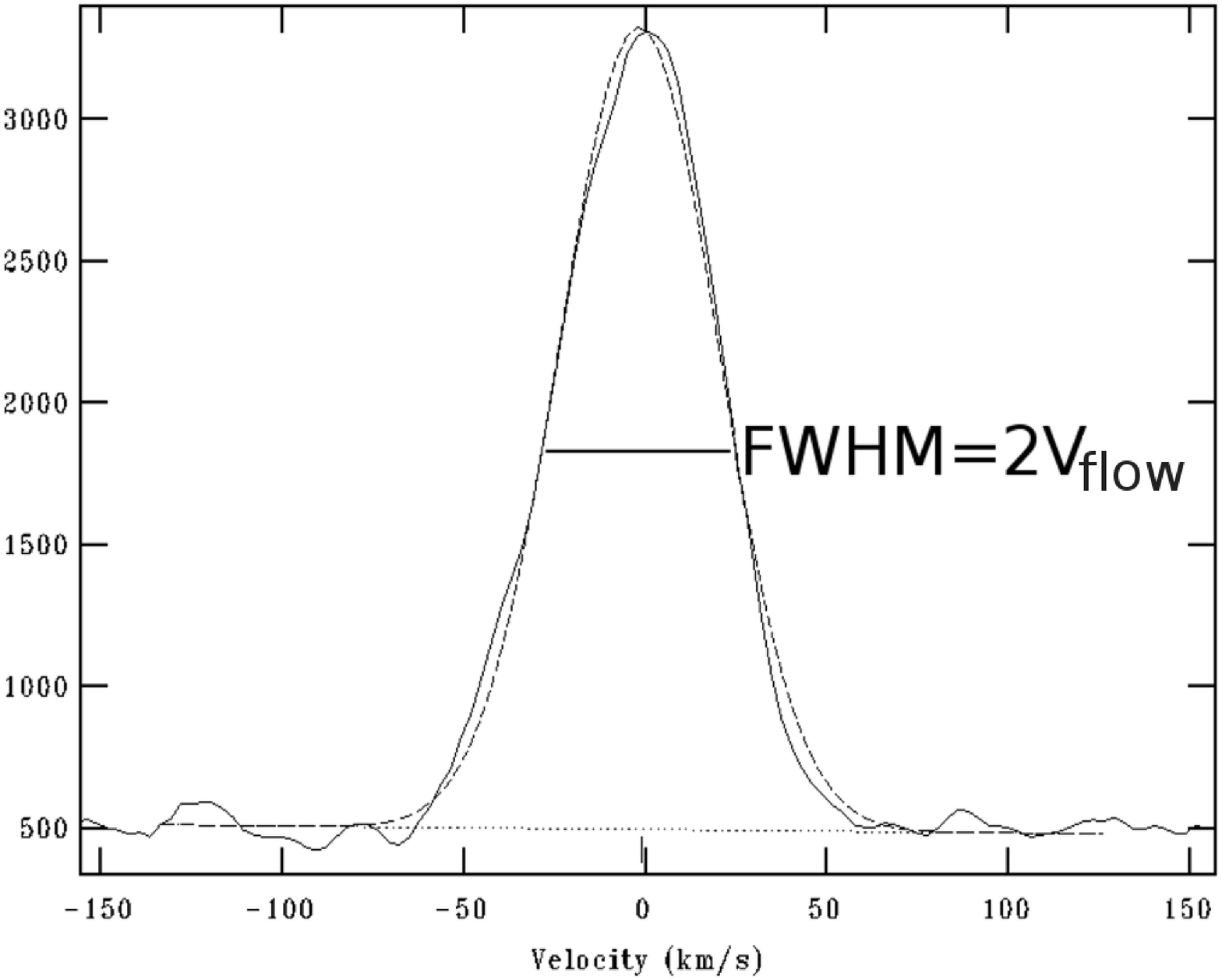} 
\caption{We present our data for NGC 246 and K1-12 to illustrate our method for determining the bulk flow velocities in the objects of our sample. In the top panel, the H$\alpha$ line profile for NGC 246 is clearly split and the bulk flow velocity is taken as half the peak to peak difference in velocity between the blue- and red-shifted components at the point of maximum splitting, at the centre of the line profile in this case. In the bottom panel, the H$\alpha$ line profile for K 1-12 is filled, so we fit a single Gaussian profile to the line profile and assign half of the resulting FWHM as the bulk flow velocity. We used the first method in the great majority of cases, including all measurements of the [\ion{N}{2}] $\lambda$6584 and [\ion{O}{3}] $\lambda$5007 lines, but the second was required for six H$\alpha$ measurements (see Table \ref{table01}).}
   \label{figure02}
\end{center}
\end{figure*}

The calibrated spectra were analyzed using the IRAF\footnote[1]{{IRAF is distributed by the National Optical Astronomy Observatories, which are operated by the Association of Universities for Research in Astronomy, Inc., under cooperative agreement with the National Science Foundation.}} package to obtain the bulk flow velocity, V$_{flow}$.  
Given the high spectral resolution and our selection of essentially round and spatially resolved objects, we expect that a spherical expansion will be represented by the shape of a typical velocity ellipse in the bi-dimensional  line profile or position - velocity ({\it P--V}) diagram (e.g., Fig. \ref{figure02}).

We measured the bulk flow velocities, $V_{flow}$, as half the peak to peak difference in velocity between the blue- and red-shifted components at the point of maximum splitting, usually the centre of the line profile.  Thus, these flow velocities correspond to the typical outflow velocities for the matter with the highest emission measure within the spectrograph slit.  Although this is often what is usually understood as the expansion velocity, in strict terms the expansion velocity should be understood as the velocity of the outer shock \citep[e.g.,][]{Schonberner05b}.  
For six objects (K1-12, A65, Kn39, A57, Pu1, PHRJ1754-3533), the H$\alpha$ line profile shows no splitting.  For these few cases, we fit a single Gaussian profile to the line profile and assign half of the resulting FWHM as the bulk flow velocity \citep{G&Z00}.  We correct the measured FWHM for instrumental broadening (11.5\,km\,s$^{-1}$ FWHM), thermal Doppler broadening (assuming a temperature of $10^4$\,K), and fine structure broadening \citep[assuming a value of 7.5 \,km\,s$^{-1}$ FWHM;][]{GD08} as in \citep{Richer08}.  
Examples of both kinds of measurement are shown in Fig. \ref{figure02}. We measured $V_{flow}$ from all available spectra (H$\alpha$, [\ion{N}{2}] $\lambda$6584, and [\ion{O}{3}] $\lambda$5007 emission lines, as may be the case).



\section{Results}\label{sec_sample_results}

\subsection{Velocity distributions}

\begin{figure}[ht]
\begin{center}
 \includegraphics [width=6.0in]{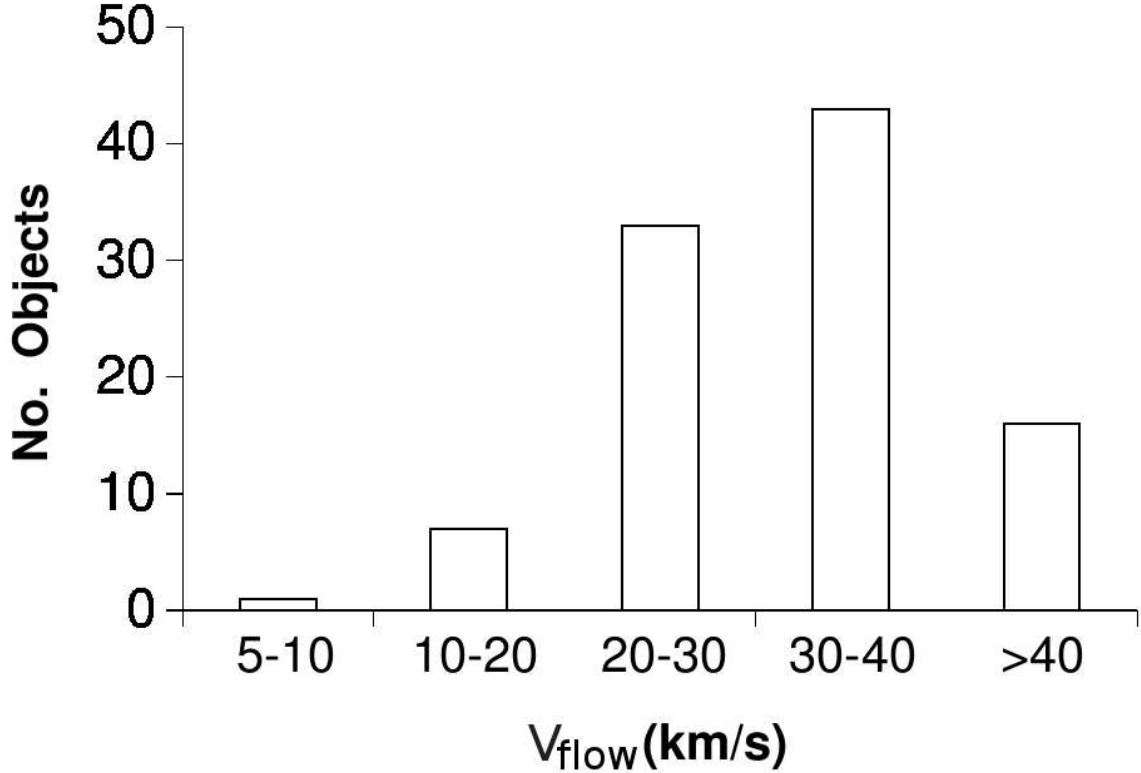} 
\caption{We present a histogram of the bulk flow velocities of the nebular shell for the evolved PNe in our sample. Here, we use velocity measurement from the [\ion{N}{2}] $\lambda$6584 line when available. Otherwise, we use the bulk flow velocity from [\ion{O}{3}] $\lambda$5007 or, as a last resort, H$\alpha$. The average V$_{flow}$ for our sample PNe (31 km s$^{-1}$ or 29 km s$^{-1}$ using V$_{H\alpha}$ only) is higher than the average V$_{flow}$ for younger PNe \citep [$\sim 25$\,km s$^{-1}$;][]{Richer08}.}
   \label{figure03}
\end{center}
\end{figure}

The histogram in Fig. \ref{figure03} shows the distribution of bulk flow velocities for our entire  sample.  Most of the PNe in this sample have bulk flow velocities exceeding 25 km s$^{-1}$, in good agreement with the predictions from theoretical models \citep{Villaver02,Perinotto04}, though there is a group of nebulae also found at lower velocities.  The bulk flow velocities in Fig. \ref{figure03} are those measured from, in order of preference, [\ion{N}{2}], [\ion{O}{3}] and H$\alpha$, depending upon the velocities available.  

\begin{figure}[ht]
\begin{center}
\includegraphics [width=6.0in]{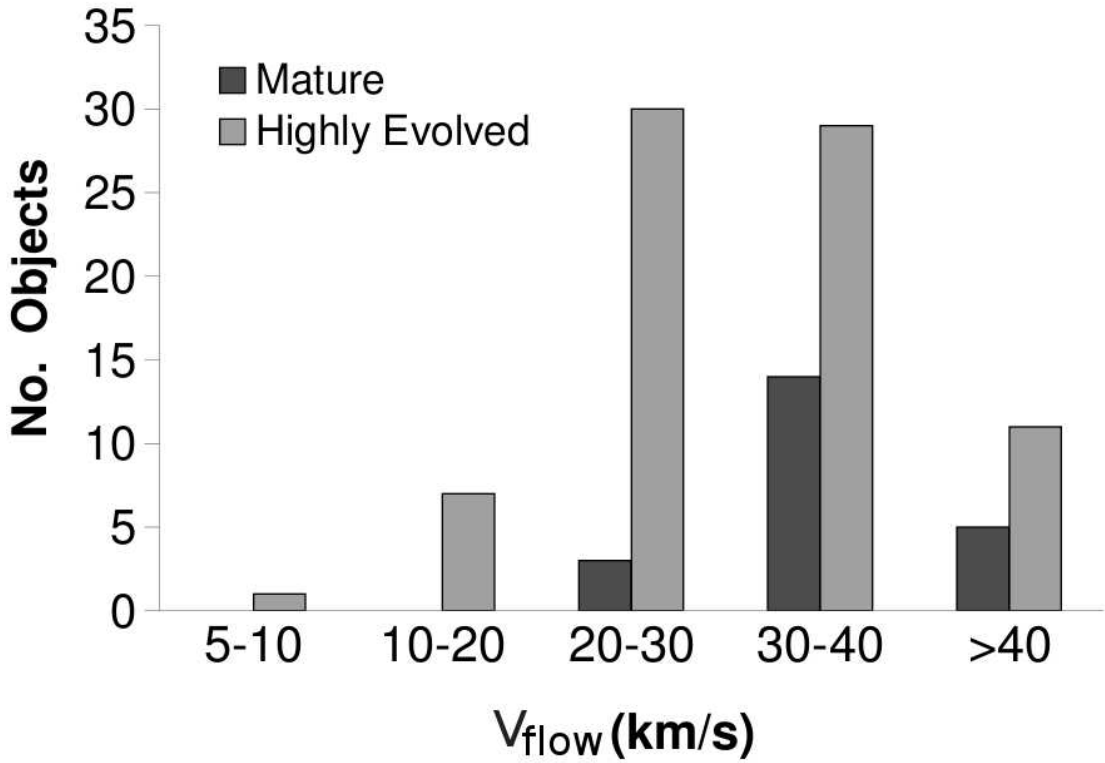} 
 \caption{These histograms present the distribution of bulk flow velocities for Mature and HE PNe. The bulk flow velocities for the 22 Mature (M) PNe are biased to larger values than those for the 78 Highly Evolved (HE) PNe.  The mean bulk flow velocity and standard deviation about the mean are 37 km\,s$^{-1}$ and 8 km\,s$^{-1}$ for Mature PNe and 30 km\,s$^{-1}$ and 9 km\,s$^{-1}$ for HE PNe.  These histograms use the same bulk flow velocities as in Fig. \ref{figure03}.} 
   \label{figure04}
\end{center}
\end{figure}

Fig. \ref{figure04} presents the  distributions of the Mature and HE PNe.  The Mature PNe preferentially populate the bins at higher bulk flow velocities. The bulk flow velocities of HE PNe span a wide range, but tend to lower values than typically found for Mature PNe \citep[see also][]{H&W90}. The median bulk flow velocities are 37 km s$^{-1}$ for Mature PNe and 30 km s$^{-1}$ for HE PNe. Based upon their CS properties and nebular spectra, HE PNe are more evolved than Mature PNe.  

\begin{figure}[ht]
\begin{center}
\includegraphics [width=6.0in]{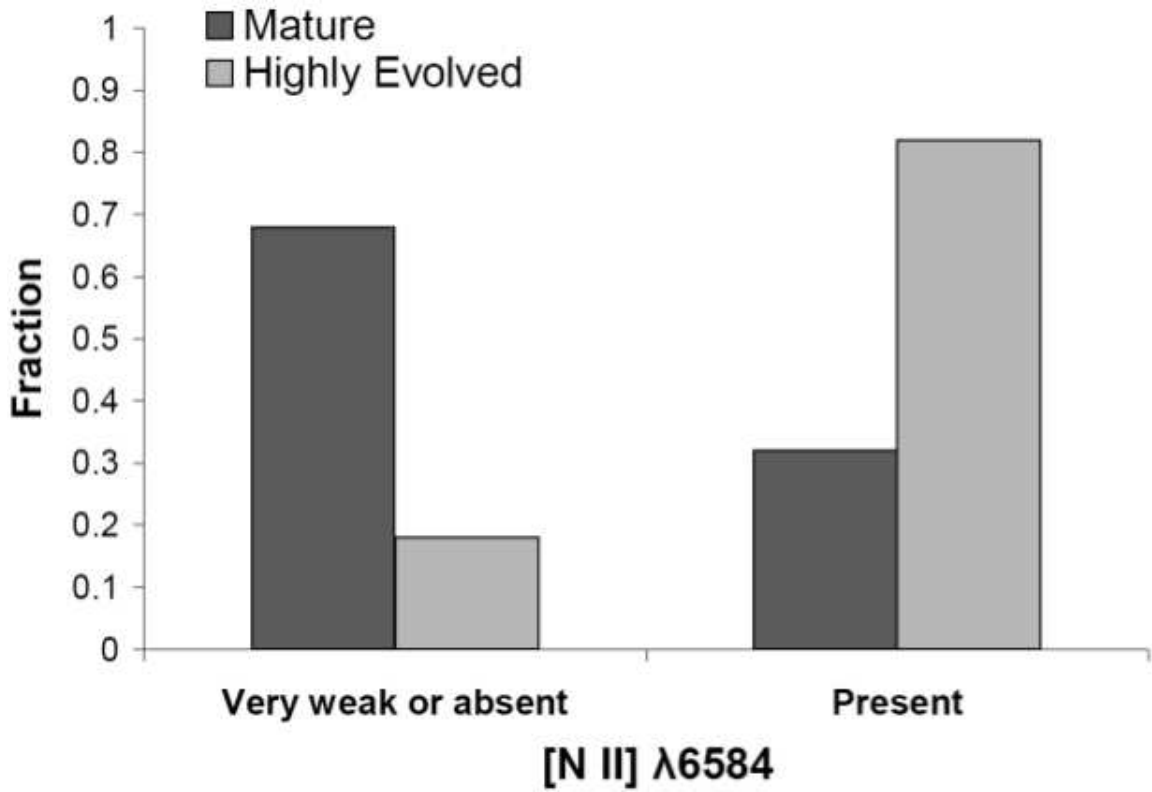} 
 \caption{The presence or absence of [\ion{N}{2}] $\lambda$6584 emission correlates with the morphological classes of Mature and HE PNe.  The Mature objects from our sample tend to have very weak or absent [\ion{N}{2}] $\lambda$6584 emission whereas the HE PNe present emision in this line.}
   \label{figure05}
\end{center}
\end{figure}
\begin{figure}[ht]
\begin{center}
\includegraphics [width=6.0in]{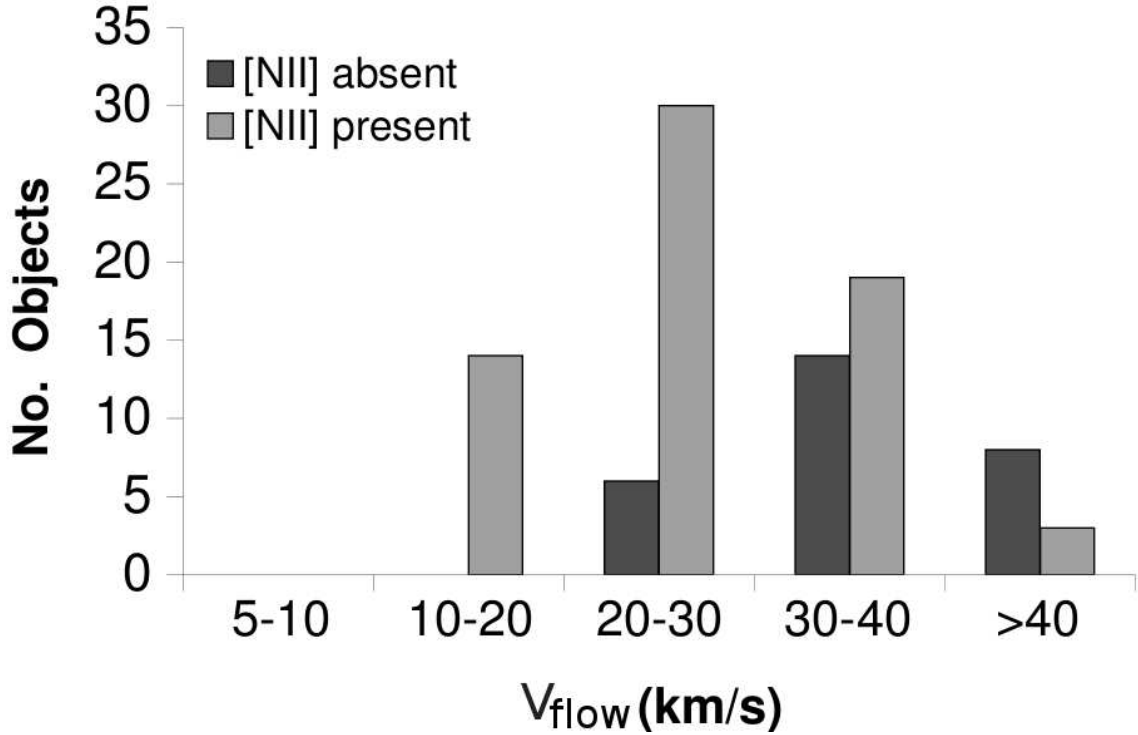} 
 \caption{The incidence of the [\ion{N}{2}] $\lambda$6584 emission line is not random as a function of the bulk flow velocity of the nebular shell, V$_{flow}$.  The absence of [\ion{N}{2}] $\lambda$6584 emission is a common characteristic of objects with high V$_{flow}$ whilst the oppossite behavior is observed for objects with low V$_{flow}$.  The similarity of this result with that of Fig. \ref{figure04} implies that our classification selects distinct evolutionary phases.  For consistency, in this plot we use velocities measured from H$\alpha$ only.}
   \label{figure06}
\end{center}
\end{figure}

The Mature PNe also distinguish themselves spectrally. 
The [\ion{N}{2}] $\lambda$6584 emission line is often absent in Mature PNe. 
From Fig. \ref{figure05}, it is clear that weak or absent [\ion{N}{2}] $\lambda$6584 emission is a common feature in the Mature PNe, but not for HE PNe, since most of the latter have [\ion{N}{2}] $\lambda$6584 emission even though they present no evidence of an ionization front.    
In Fig. \ref{figure06}, 
objects with weak or absent [\ion{N}{2}] $\lambda$6584 emission have higher bulk flow velocities than do PNe who present this emission.  The similarity of Figs. \ref{figure04} and \ref{figure06} indicates that our rough classification indeed distinguishes two different evolutionary stages.  

The spectral difference between HE and Mature PNe found in Fig. \ref{figure05} introduces a systematic bias into Fig. \ref{figure04}.  Typically, the [\ion{N}{2}] velocities are usually slightly larger than the [\ion{O}{3}], and H$\alpha$ velocities (the latter two are similar; c.f. Table \ref{table01}).  On average, then, the bulk flow velocities used for HE PNe in Fig. \ref{figure04} will be slightly over-estimated compared to those used for Mature PNe.  This bias acts to \emph{minimize} the difference between the two samples, so, in spite of this bias, the difference between the two samples persists and does not affect the basic result (see \S \ref{section_tests}, Fig. \ref{figure10}, panel (c)).

\subsection{Location in the H-R diagram}\label{sec_subsample}

From the discussion in \S \ref{sec_introduction}, we expect the CSs in our sample of PNe to be found on or near the white dwarf cooling track in the H-R diagram.  
We searched for luminosities and temperatures for the CSs in our sample to locate them in a H-R diagram 
\citep[e.g.,][among many others]{Kaler90, STG02, Frew08}.  To avoid the uncertanties due to different distance scales (different systematic errors), we adopted the data based upon the \citet{Frew08} distance scale.  Our primary reasons for doing so are that this distance scale was designed for evolved objects, there exist data for a large number of our objects, and the concordance with theoretical evolutionary tracks is reasonably good \citep{VW94, Blocker96, STG00}.  

\begin{figure}[ht]
\begin{center}
 \includegraphics [width=6.0in]{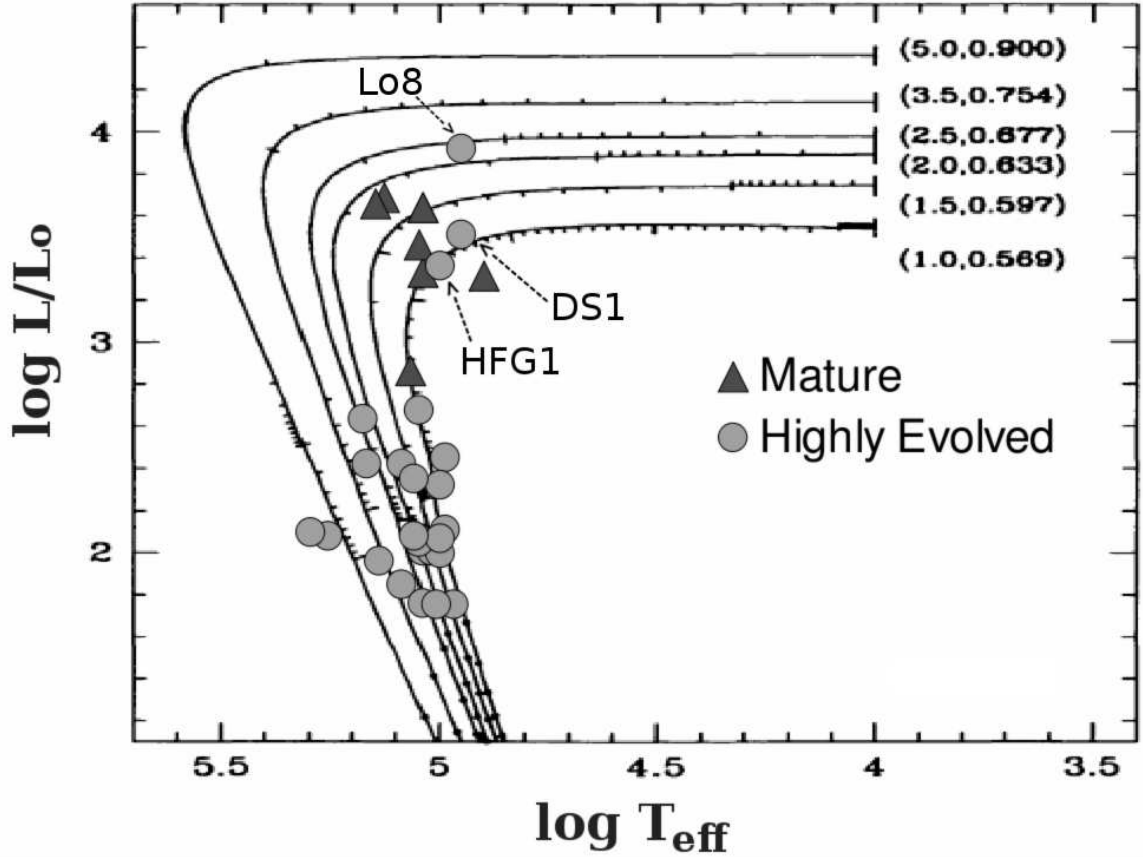} 
 \caption{
The locations of Mature and HE PNe in the H-R diagram, with evolutionary tracks from \cite{VW94}, reveal a systematic difference, again reflecting different evolutionary stages.  Mature PNe (8 PNe) have systematically higher luminosities than HE PNe (24 PNe), which are already evolving towards to white dwarf phase.  DS 1 and HFG 1 contain binary central stars whose luminosities are uncertain.}
   \label{figure07}
\end{center}
\end{figure}

In Fig. \ref{figure07} we present the H-R diagram of our objects with available data, with different symbols for our two classes (Mature or HE PNe).  Again, there is a clear separation, indicating that the two classes represent distinct evolutionary stages.  
Mature PNe are located near the point of maximum temperature of the evolutionary tracks.  These objects are therefore at a moderately evolved stage and have high luminosity CSs.  On the other hand, the vast majority of HE PNe are well down the white dwarf cooling track and have low luminosity CSs. 

\begin{figure}[ht]
\begin{center}
 \includegraphics [width=6.0in]{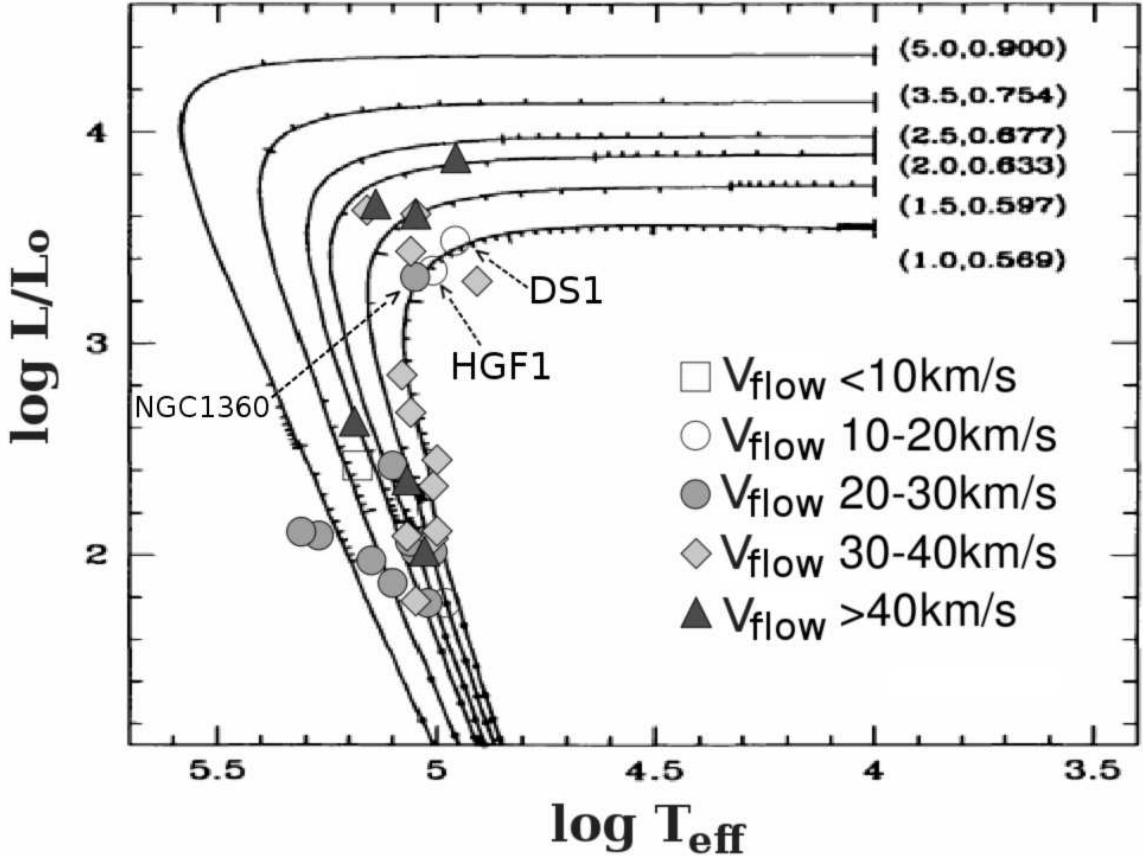} 
 \caption{We find a clear evolution of the kinematics for the most evolved phases of evolution of the nebular shell.   The nebular shell decelerates systematically as the central star luminosity decreases.  
The mean bulk flow velocity and standard deviation about the mean are 37 km\,s$^{-1}$ and 4 km\,s$^{-1}$ for Mature PNe (8 PNe) and 28 km\,s$^{-1}$ and 10 km\,s$^{-1}$ for HE PNe (24 PNe), similar to what are found for the entire sample (Fig. \ref{figure04}).  Here, we use the same bulk flow velocities 
as in Fig. \ref{figure03}. Although there is some mixture of velocities at high and low luminosities, we find 
a clear decrease in the V$_{flow}$ as the CS luminosity drops. Again, the evolutionary tracks are from \citep{VW94}.  DS 1 and HFG 1 contain binary central stars whose luminosities are uncertain.}
  \label{figure08}
\end{center}
\end{figure}

In Fig. \ref{figure08}, we present the same data, but now using different symbols for different ranges of bulk flow velocities. Again, there is a clear general trend for the evolution of the kinematics of the nebular shell, with the highest bulk flow velocities found at the highest luminosities and the lowest velocities at the lowest luminosities. 
In general, as the CS luminosity declines, we observe a deceleration of the nebular shell, as predicted by some hydrodynamical models \citep{Villaver02,Perinotto04}.  Since we use the same bulk flow velocities as in Figs. \ref{figure03} and \ref{figure04}, Fig. \ref{figure08} incorporates the same bias as Fig. \ref{figure04} in the same sense.  Clearly, it does not affect our conclusion (see \S \ref{section_tests}).

\begin{figure}[ht]
\begin{center}
\includegraphics [width=4.0in]{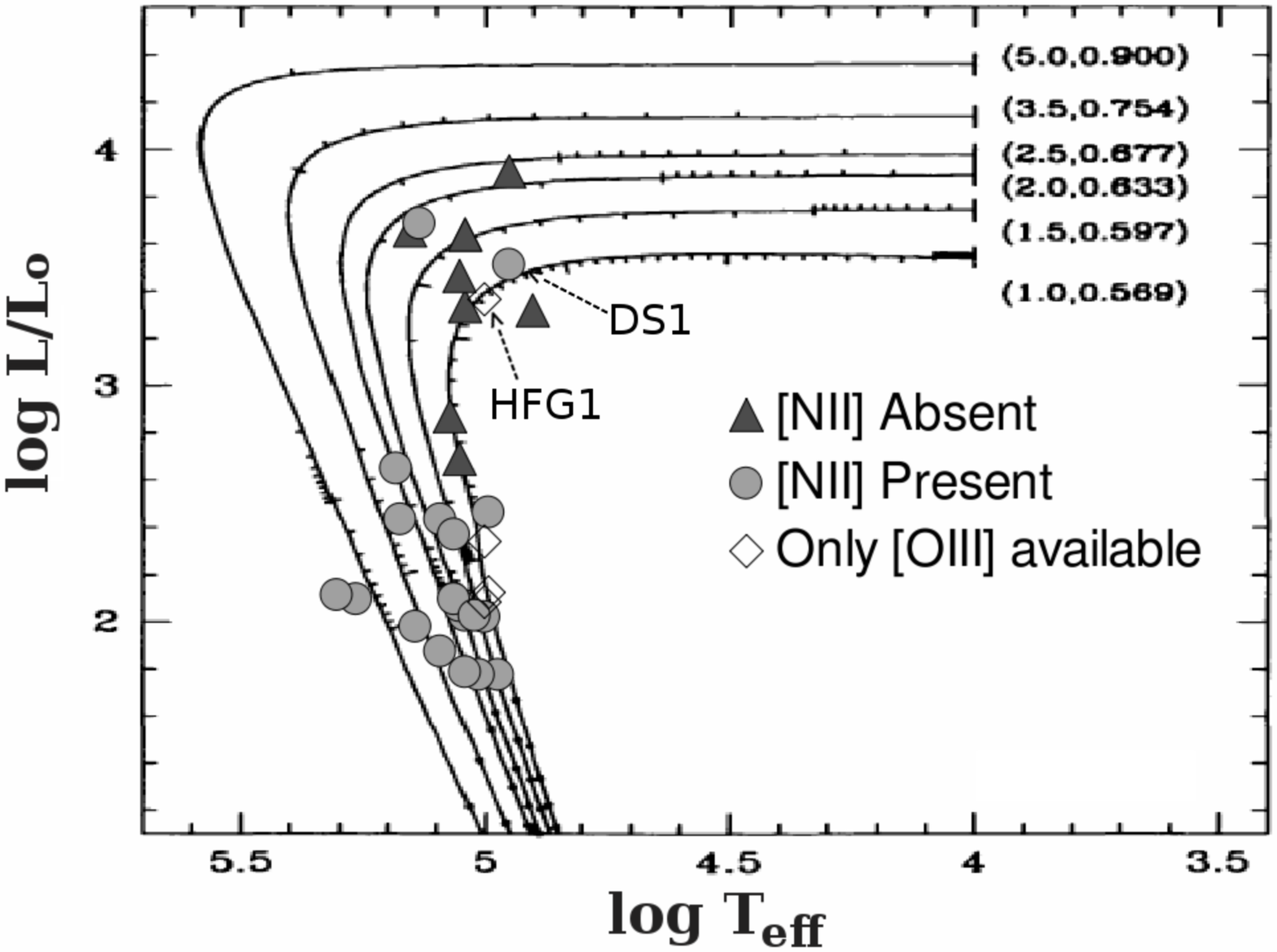}
\includegraphics [width=4.0in]{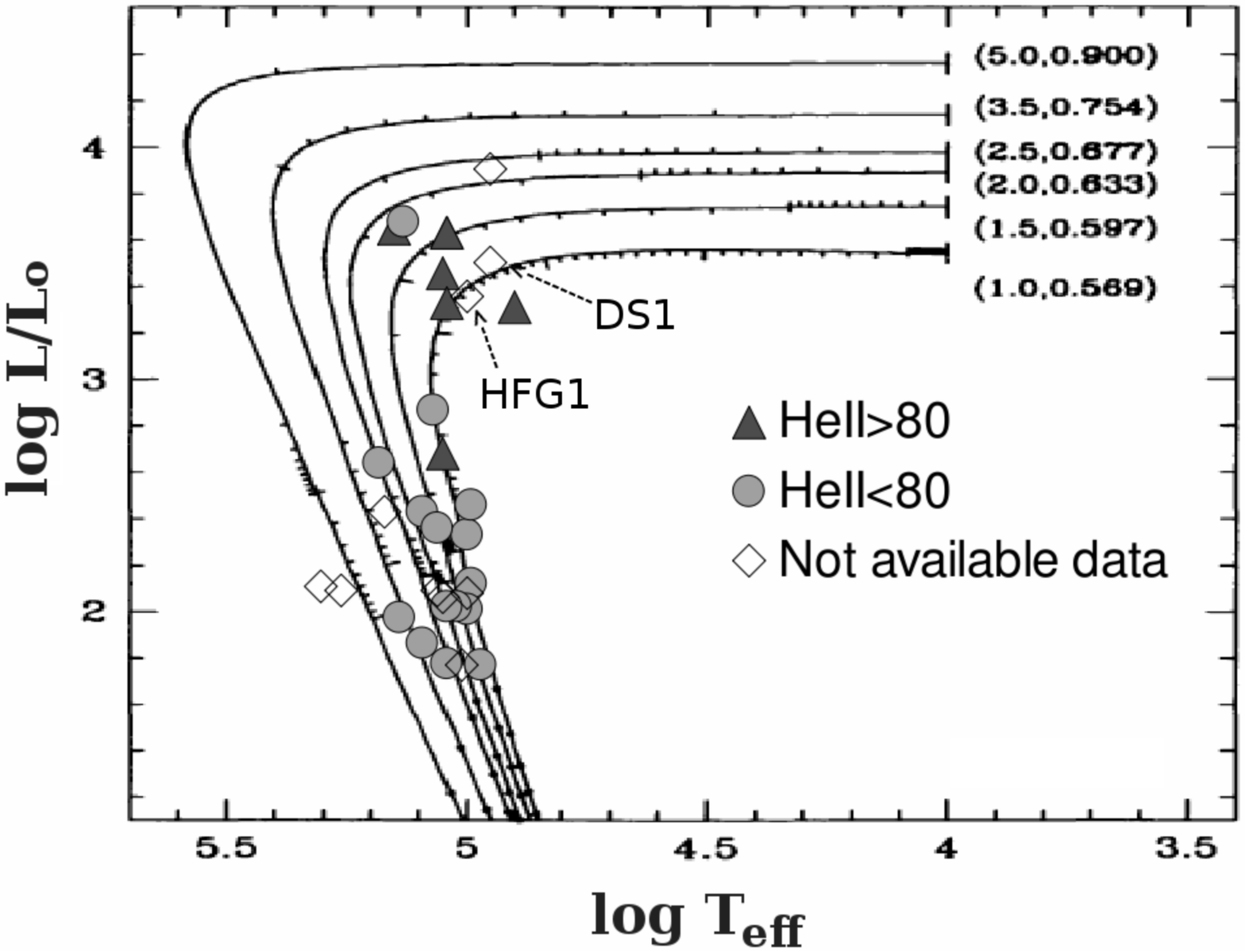} 
\caption{The degree of excitation of the nebular spectrum supports the luminosities deduced for the CSs in our sample.  Since the CS temperature varies little, the degree of ionization is controlled by the CS luminosity.  Here, we see that nebulae surrounding high luminosity CSs typically have weak or absent [\ion{N}{2}] $\lambda$6584 emission, but strong \ion{He}{2} $\lambda$4686 emission, reflecting their high degree of excitation.  The opposite occurs for the nebulae surrounding low luminosity CSs.  Considered with Figs. \ref{figure07} and \ref{figure08}, it is evident that the CS luminosity, nebular morphology, nebular bulk flow velocity, and the degree of nebular excitation vary coherently and consistently.  Once more, the evolutionary tracks are from \citep{VW94}.  DS 1 and HFG 1 contain binary central stars whose luminosities are uncertain.  
}
   \label{figure09}
\end{center}
\end{figure}

In Fig. \ref{figure09}, we present the line intensity data for our sample of PNe.  The top panel concerns the presence or absence of the [\ion{N}{2}] $\lambda$6584 emission line.  Most of the objects where this line is weak or absent are near the point of maximum temperature on the evolutionary tracks.  The bottom panel in Fig. \ref{figure09} considers the intensity of \ion{He}{2} $\lambda$4686 \citep[$I_{\mathrm{He\,\sc{II}}}$;][]{Tylenda94}, where we divide our objects into groups of higher and lower degrees of excitation.  For the higher degree of excitation, we adopt $I_{\mathrm{He\,\sc{II}}}>80$ (scale: $I_{\mathrm H\beta}=100$), with the remaining objects being of lower degree of excitation.  (By normal standards, all of these objects are of a high degree of excitation.) In this plot, the objects with the higher degree of excitation tend to have higher luminosities.  There is little variation in temperature for the CSs of these evolved PNe, so the degree of excitation is controlled principally 
by the luminosity of the CS.  Where \ion{He}{2} $\lambda$4686 is strongest, [\ion{N}{2}] $\lambda$6584 emission is generally weak or absent due to the high degree of ionization.  

From Figs. \ref{figure07}-\ref{figure09}, the PNe with high luminosity central stars typically tend to (i) fall into our mature PN category, (ii) lack the presence of [\ion{N}{2}] $\lambda$6584 emission, (iii) have the strongest \ion{He}{2} $\lambda$4686 lines, and (iv) have higher bulk flow velocities.  While we shall see that the above hold true generally in a statistical sense (\S \ref{section_tests}), they are not strict general rules and exceptions to each are observed.  An obvious complication in these diagrams is the effect of binary central stars, and, indeed, DS1 and HFG1, two of the anomalous data points in all of these diagrams, appear to have binary central stars \citep{Montez10}.  Fortunately, our large data set allows us to distinguish these general trends in spite of these complications.  

\subsection{Statistical tests}\label{section_tests}

\begin{figure}[ht]
\begin{center}
\includegraphics [width=3.0in]{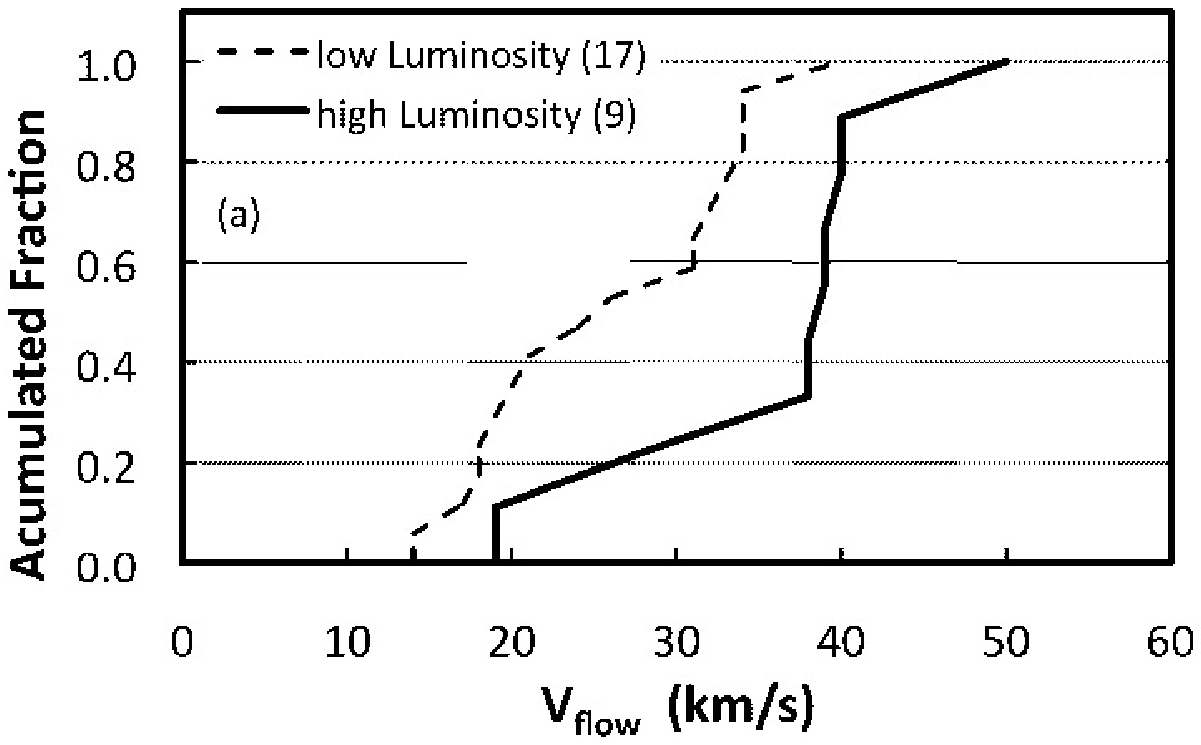}
\includegraphics [width=3.0in]{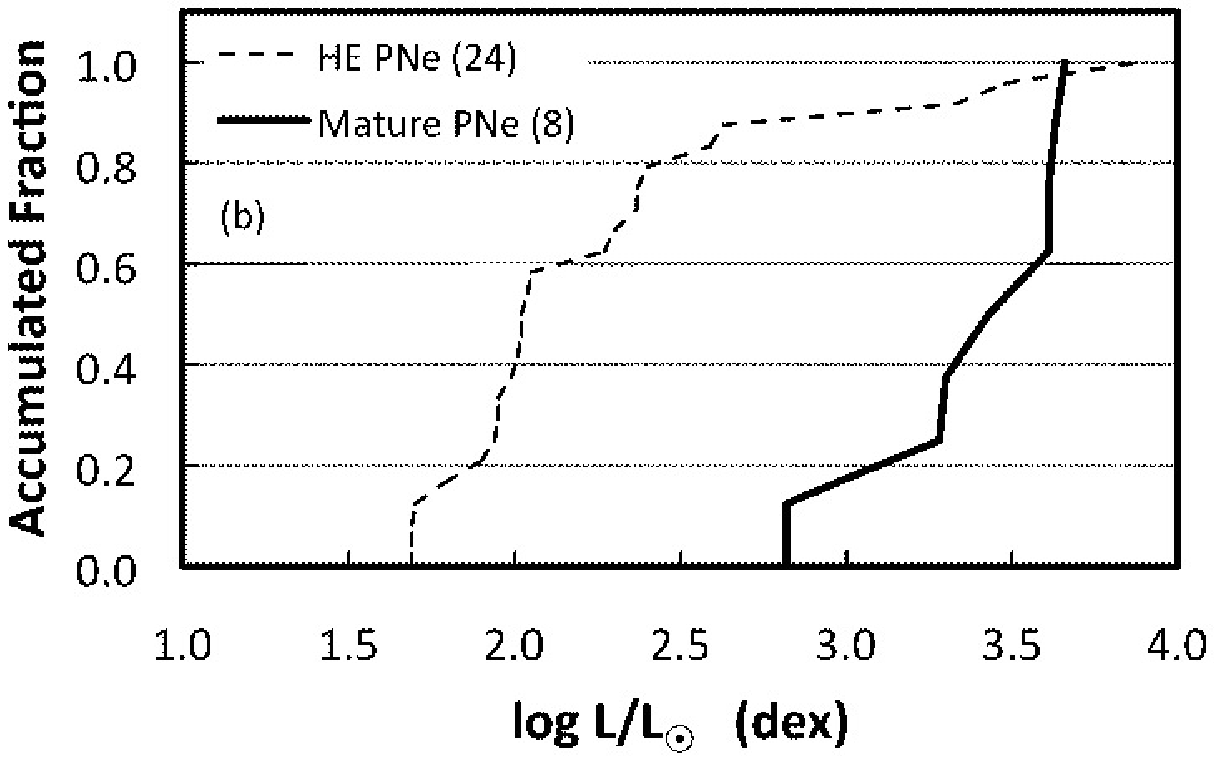} 
\includegraphics [width=3.0in]{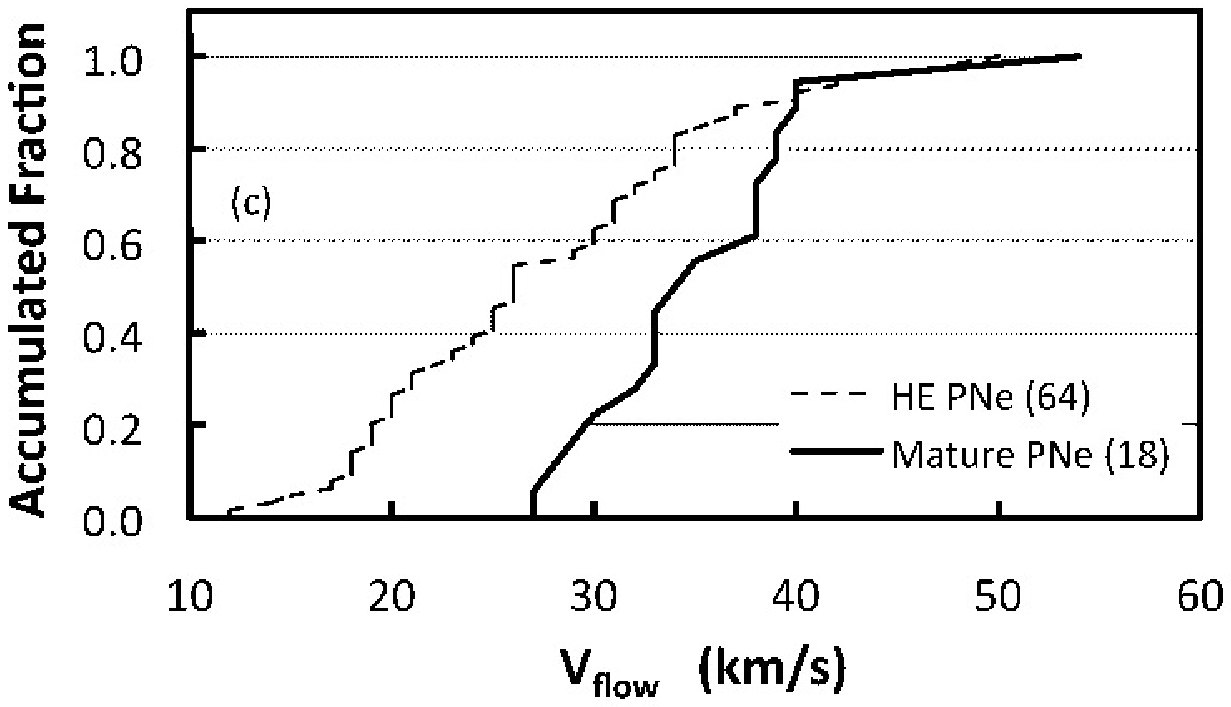}
\includegraphics [width=3.0in]{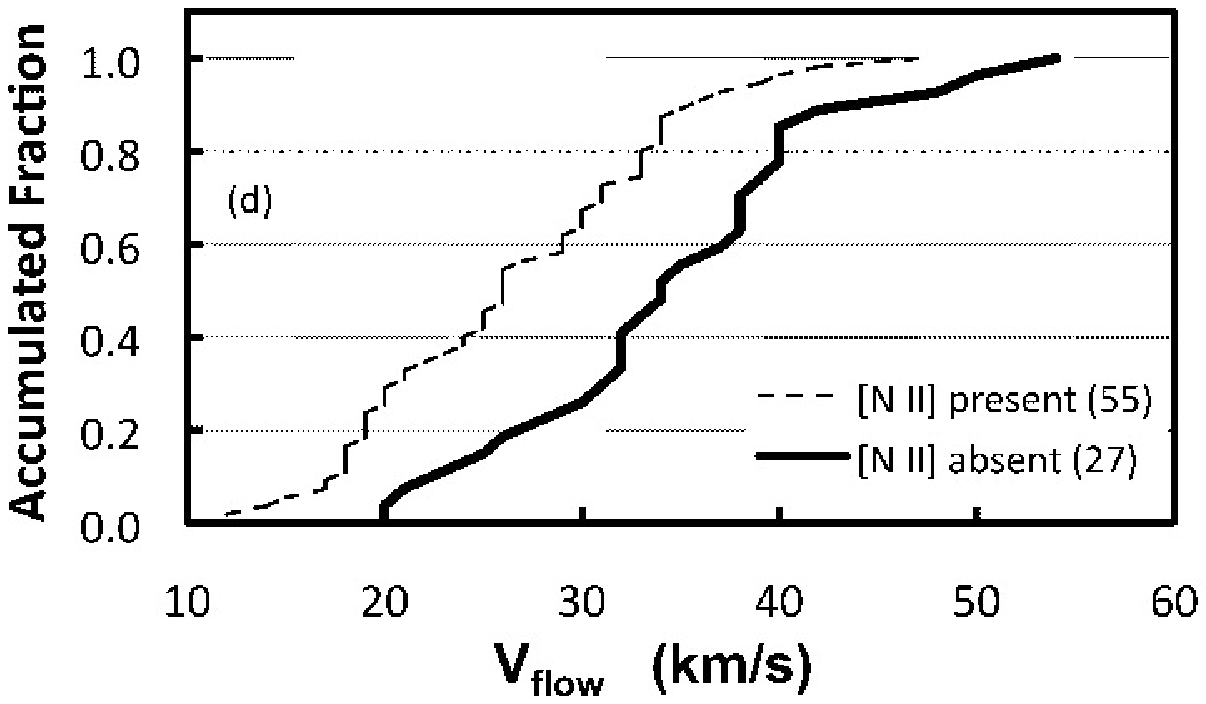}
\caption{We present the cumulative distribution of $V_{flow}$ or CS luminosity for different subsamples of our objects that we used to test (a) whether PNe surrounding high and low luminosity CS have the same bulk flow velocities, whether Mature and HE PNe have (b) similar CS luminosities or (c) bulk flow velocities, and (d) whether PNe with and without [\ion{N}{2}] $\lambda$6584 emission have similar bulk flow velocities.  In all cases, the distributions differ at high statistical significance, based upon the KS and U tests (see text for details).  For these tests, we use $V_{flow}$ measured from the splitting of the H$\alpha$ line (method \#1, Fig. \ref{figure02}).  The numbers in parentheses indicate the number of objects in each sample.}
   \label{figure10}
\end{center}
\end{figure}

In Fig. \ref{figure10}, we present cumulative distributions to quantify the foregoing statistically.  For the distributions in panels (b), (c), and (d), we use $V_{flow}$ measured from the splitting of the H$\alpha$ line.  The results do not depend significantly upon this choice of data and we obtain similar results using the data adopted for Figs. \ref{figure03} and \ref{figure04}.  However, the comparison among objects is clearer if we adopt a single emission line and measurement method for all objects.  The data set used in panels (a) and (b) is entirely drawn from the \citet{Frew08} sample, which is volume-limited.  

In panel (a), we plot the cumulative distributions of the bulk flow velocities for PNe with CSs of high and low luminosity.  We divide the central star luminosities at a luminosity of $L/L_{\odot} = 3.0$ dex because (1) it approximately divides between Mature and HE PNe (Fig. \ref{figure07}), (2) it roughly divides the PNe between those of higher and lower nebular excitation (Fig. \ref{figure09}), and (3) theoretically, we expect a gap in luminosity at about this value because of the rapid decrease in luminosity following the termination of nuclear reactions.  The accumulated fraction is the fraction of objects in a given sample with bulk flow velocities below the value plotted, e.g., $$\mathrm{Accumulated\ Fraction}(V_{flow})= \frac{N(V < V_{flow})}{N_{total}},$$ where $V_{flow}$ is the abscissa value, $N(V < V_{flow})$ is the number of objects with bulk flow velocities less than $V_{flow}$, and $N_{total}$ is the total number of objects in the sample.  Thus, from the first panel in Fig. \ref{figure10} we see that none of  the high luminosity objects have bulk flow velocities below 18 km\,s$^{-1}$ and the majority have bulk flow velocities exceeding 37 km\,s$^{-1}$ whereas 80\% of the low luminosity objects have bulk flow velocities below 37 km\,s$^{-1}$.  

Two common tests for comparing such cumulative distributions are the Kolmogorov-Smirnov (KS) test and the Wilcoxon-Mann-Whitney non-parametric U-test \citep{wallandjenkins2003}.  The KS test uses the sample sizes and the maximum difference between the ordinate values of the two distributions to determine the probability that the two distributions may be drawn randomly from a single parent distribution.  The U test is conceptually similar to the Student's t test \citep{wallandjenkins2003} in that it uses the rank values of the data from the two samples to derive statistics analogous to the mean and variance and uses these to determine the probability that the two samples may be drawn randomly from the same parent distribution.  Using the KS and U tests to compare the  distributions for PNe with high- and low-luminosity CSs in the first panel in Fig. \ref{figure10}, we find a 99\% probability that they do not arise from the same parent distribution.  Note that we do not consider whether a given PN falls in the Mature or HE category for this comparison, but we can only use the data for PNe with know CS parameters.  

In panel (b) of Fig. \ref{figure10}, we present the cumulative distributions of the CS luminosity for Mature and HE PNe, finding extremely small probabilities ($\ll 1$\%) from both the KS and U tests that the two distributions arise from the same parent distribution.  Clearly, HE PNe have lower luminosity CSs.  Again, this analysis includes only the objects in our sample with known CS parameters.  

In panels (c) and (d) of Fig. \ref{figure10}, we analyze our entire PN sample.  Panel (c) presents the cumulative distributions of bulk flow velocities for Mature and HE PNe.  Panel (d) presents the cumulative distributions of bulk flow velocities for objects in which the [\ion{N}{2}] $\lambda$6584 emission line is present and absent.  In both cases, both the KS and U tests return very small probabilities ($< 1$\%) that the distributions arise as a result of a random draw from the same parent distribution.  We can conclude that HE PNe have lower bulk flow velocities than Mature PNe and that PNe in which the [\ion{N}{2}] $\lambda$6584 is present likewise have lower bulk flow velocities than those in which this line is absent.  If we restrict these tests to the subset of objects from \citet[][]{Frew08} as in panels (a) and (b), we obtain similar results, though the significance for the K-S test is 96\% and 98\% for panels (c) and (d), respectively, while the significance from the U test remains $\ll 1\%$.

Therefore, our statistical analysis reveals that, with high significance, (i) nebulae surrounding CSs of high luminosity expand faster than do those surrounding low luminosity CSs, (ii) Mature PNe have higher luminosity CSs than do HE PNe, (iii) Mature PNe have larger bulk flow velocities than HE PNe, and (iv) PNe in which the [\ion{N}{2}] $\lambda$6584 line is absent have larger bulk flow velocities than do their counterparts for which this line is present.  Since CS luminosity, nebular morphology, and nebular excitation are all related to the evolutionary stage of the PN system, our results indicate clear trends with evolutionary stage.  High luminosity CSs are found in PNe with higher nebular excitation, better-defined morphological structure, and higher bulk flow velocity.  PNe with low luminosity CSs have lower excitation, more diffuse structure, and lower bulk flow velocity.  

\section{DISCUSSION}\label{sec_discussion}

We have caracterized the global kinematics of the nebular shell in PNe at advanced stages of evolution with the largest sample used to date for this purpose.  Generally, we find a good correlation between the evolutionary stage of the CS and the kinematics, morphology, and spectral properties of the nebular shell, as predicted by modern hydrodynamical models \citep[e.g.,][]{Villaver02, Perinotto04, Schonberner05b, Schonberner10}.
Nebular shells with the highest degrees of excitation, largest bulk flow velocities, and best-defined structure usually have moderately evolved, high luminosity CSs that inhibit [\ion{N}{2}] $\lambda$6584 emission.  On the contrary, nebular shells with lower degree of excitation, lower bulk flow velocity, and the least structure contain 
the most evolved CSs.  In agreement with \citet{Richer10}, we find that the highest bulk flow velocities occur for CSs with temperatures and luminosities for which we expect 
the stellar wind still has a strong influence on the kinematics of the nebular shell.  At the latest stages of PN evolution, we find clear observational and statistical evidence for 
the deceleration of the nebular shell. 

While it is clear that the nebular shell decelerates with time, its interpretation is not necessarily so simple.  As the CS ceases nuclear reactions at the point of maximum temperature, its luminosity drops, initially precipitously, then more slowly.  At the same time, the (luminosity-driven) wind energy imparted to the hot bubble likewise drops.  However, the over-pressure of the hot bubble is likely to continue to affect the kinematics of the nebular shell until the expansion of the hot bubble causes its pressure to drop significantly.  At some point, the pressure must drop, decelerating the expansion of the nebular shell, but it might be so severe to allow some backflow of the nebular shell towards the central star \citep[e.g.,][]{GS06}.  In addition, the loss of CS luminosity may allow recombination of the nebular shell.

The CS luminosities of the Mature and HE PNe differ by an order of magnitude for the objects in our sample (Figs. \ref{figure07}-\ref{figure09}).  Thus, it is unclear whether the HE PNe are entirely ionized and that their lower luminosity CSs can maintain the same mass ionized as do the higher luminosity CSs in Mature PNe.  If not, recombination will occur in the outermost part of the nebular shell and, in HE PNe, we may observe only the inner-most material observed in Mature PNe.  If that is the case, the lower bulk flow velocities in HE PNe could reflect the slower motions of matter closer to the CS at earlier evolutionary stages.  
  
It is also conceivable that the CS is not the only or the most important influence at the latest evolutionary stages.  In some instances, interaction with the ambient interstellar medium (ISM) may affect the nebular kinematics.  \citet{oeygarciasegura2004} find that this effect is important for the evolution of ISM superbubbles and \citet{villaveretal2002a} find that it is responsible for slowing down the AGB wind in their models.  In our sample, we find that all PNe, whether classified as Mature or HE, more than $\sim 350$\,pc from the galactic plane expand more rapidly than 30\,km s$^{-1}$ \citep[see also][]{H&W90}.  However, this may not constitute evidence that the ambient ISM decelerates HE PNe, since we cannot guarantee that our sample is statistically representative within or beyond this scale height.



Any study of nebular sizes immediately confronts the uncertainties in the distances, which are considerable.  Plotting the nebular size as a function of the CS luminosity or the bulk flow velocity of the nebular shell, we do not find that HE PNe (or any substantial sub-group of these) have smaller radii than Mature PNe (Table \ref{table01}).  Neither plot shows an especially tight correlation, though that between nebular radius and CS luminosity shows more, perhaps because both depend upon the distance.  On average, then, if the loss of CS luminosity allows recombination, it is either not sufficient to cause a decrease in the nebular size or is sufficiently short that our HE PNe have all been re-ionized \citep[see][Fig. 21]{Perinotto04}.


Further observational progress is possible, but requires efforts on several fronts.  First, more reliable distance scales are necessary.  Second, in order to apply these distance scales, larger databases of homogeneous data are necessary, for both nebular shells and central stars.  Third, more detailed knowledge of central stars is necessary, particularly concerning binarity.  For example, DS1 and HFG1 are among the \lq\lq anomalous" data points in Figs. \ref{figure07}-\ref{figure09} whose CS luminosity disagrees with the nebular morphology, bulk flow velocity, and degree of excitation. For these two cases, \citet{Montez10} found X-ray evidence for binarity that could contribute to a mistaken estimate of the luminosity data from \cite{Frew08}. In fact, if we consider the luminosity values presented by \cite{Montez10} for the whole binary system, the obtained luminosity is less than the value estimated by Frew for both PNe. 
On the other hand, we cannot avoid the presence of some objects in the sample that behave in a peculiar way and for which it is not easy to find an explanation.  Lo 8, for example, has spectral features of a mature PN, high luminosity, and even evidence for the continued presence of a stellar wind \citep{Patriarchi91}, but clearly exhibits a HE PNe morphology.  Its discrepant location in the H-R diagram is indicated in Figure 7.  Another case is the well studied PN NGC 1360, an object with an accurate luminosity estimate, but whose bulk flow velocity (26 km s$^{-1}$) does not agree with the expectations for its evolutionary stage 
\citep[see Fig. \ref{figure08}; for contrasting discussions:][]{GD08, Schonberner10}. Nonetheless, the use of large samples such as ours allows us to distinguish the general trends in the data and recognize individual discrepancies for particular objects that deserve further study.

\section{Conclusions}\label{sec_conclusions}

We have analyzed the kinematics of evolved PNe using the largest homogeneous data set to date, drawn from the SPM Catalogue.  We characterize the kinematics using the bulk flow velocities for these PNe, i.e., the typical outflow velocity for the matter with the highest emission measure within the spectrograph slit.
Typically, the bulk flow velocities are larger than those for younger PNe, i.e., PNe whose CSs are on the horizontal part of the evolutionary track in the H-R diagram \citep[e.g.,][]{Richer08, Richer10}.  
We find a clear kinematic evolution of the nebular shell at these most advanced evolutionary stages, correlating with the CS evolutionary stage as well as the nebular morphology and the degree of excitation of the nebular spectrum.  As expected from hydrodynamical models \citep{Villaver02, Perinotto04}, Mature PNe, whose central stars are still at high luminosity and whose hot bubbles are still adequately powered, have more structured nebular morphologies, larger bulk flow velocities, and nebular spectra with a higher degree of excitation.  Highly evolved PNe have low luminosity CSs, now on the white dwarf cooling tracks, and nebular shells whose morphology have the least structure, lower bulk flow velocities, and spectra with a lower degree of excitation.  

These results complement prior work, illustrating the coupling between the CS's evolution and that of the nebular shell predicted by theory \citep[e.g.,][]{Villaver02, Perinotto04}.  The bulk of the mass of the nebular shell is continuously accelerated until the CS reaches its maximum temperature and ceases nuclear burning \citep[e.g.,][]{Richer10}.  Thereafter, the bulk of the mass of the shell decelerates as the CS approaches the white dwarf stage of evolution, presumably due to the loss of energy from the central star or possibly its interaction with the surrounding ISM or even due to recombination of the outermost nebular matter.  Distinguishing among these options is not simple and will be done best on a case-by-case basis.


\acknowledgments
The authors gratefully acknowledge financial support from CONACyT (82066) and UNAM-PAPIIT (IN116908, IN110011) grants. M.P. is also grateful to the Direcci\'on General de Estudios de Posgrado de la UNAM for additional financial assistence.  We thank the anonymous referee for helpful comments.

\begin{deluxetable}{llllcccccccccc}
\tabletypesize{\scriptsize}
\rotate
\tablecaption{Evolved Planetary Nebula Sample\label{table01}} 
\tablewidth{0pt}
\tablehead{
\colhead{Object}&\colhead{PN G}& \colhead{V$_{[N II]}$}&\colhead{V$_{H\alpha}$}&\colhead{V$_{[O III]}$}&\colhead{[N II]}&\colhead{Group$^a$}&\colhead{log L/Lo $^b$}&\colhead{log T $^b$}&\colhead{I(HeII) $^c$}\\
\colhead{}&\colhead{}& \colhead{$\pm2$(km s$^{-1}$)}&\colhead{$\pm2$(km s$^{-1}$)}&\colhead{$\pm2$(km s$^{-1}$)}&\colhead{Present}&\colhead{}&\colhead{}&\colhead{K}&\colhead{H$\beta$=100}\\
}
\startdata
A 01$^1$ & 119.4+06.5 &          41 &         34 &            &        yes &         HE &            &            &            \\
A 03 & 131.5+02.6 &           33 &         24 &            &        yes &         HE  &            &            &         65 \\
A 06$^1$ & 136.1+04.9 &      26 &         30 &            &        yes &         HE &              &            &          7 \\
A 08 & 167.0-00.9 &         37 &         36 &            &        yes &         HE  &            &            &         30 \\
A 13$^1$ & 204.0-08.5&      24 &         20 &            &        yes &         HE &      1.99 &       5.05 &            \\
A 15 & 233.5-16.3 &            &         25 &            &         no &         HE &           &            &        130 \\
A 16 & 153.7+22.8  &           &            &         37 &     [OIII] &         HE &            &            &         33 \\
A 18 & 216.0-00.2 &        15 &         12 &            &        yes &         HE  &            &            &            \\
A 19 & 200.7+08.4 &        21 &         19 &            &        yes &         HE &              &            &            \\
A 20 & 214.9+07.8  &          &         33 &            &         no &          M &               &            &        150 \\
A 21 & 205.1+14.2 &        29 &         26 &            &        yes &         HE &          2.04 &        5.3 &            \\
A 24 & 217.1+14.7 &        20 &         19 &            &        yes &         HE &            1.9 &       5.14 &         33 \\
A 26 & 250.3+00.1 &         27 &         17 &            &        yes &         HE  &               &            &            \\
A 29 & 244.5+12.5 &         23 &            &            &        yes &         HE &             1.69 &       5.01 &            \\
A 31 & 219.1+31.2 &        18 &         17 &            &        yes &         HE &             1.69 &       4.97 &         30 \\
A 33 & 238.0+34.8  &           &            &         37 &     [OIII] &         HE &            2.27 &          5 &         58 \\
A 34 & 248.7+29.5  &           &            &         34 &     [OIII] &         HE &         2.05 &       4.99 &         40 \\
A 36 & 318.4+41.4  &           &         38 &            &         no &          M &             3.43 &       5.05 &        118 \\
A 39 & 047.0+42.4  &            &         32 &            &         no &          M &        2.82 &       5.07 &         69 \\
A 40 & 359.1+15.1 &            &         35 &            &         no &          M &                 &            &            \\
A 42 & 016.0+13.5  &           &         38 &            &         no &          M &             &            &            \\
A 43 & 036.0+17.6  &           &         40 &            &         no &          M &        3.61 &       5.04 &         93 \\
A 45& 020.2-00.6 &          20 &         21 &            &        yes &         HE &        2.02 &       5.26 &            \\
A 50 & 078.5+18.7 &         33 &         30 &            &        yes &         HE &            &            &         34 \\
A 57$^2$ & 058.6+06.1 &        &       24 * &            &        yes &          M &            &            &            \\
A 65$^2$ & 017.3-21.9 &     28 &       26 * &            &        yes &         HE &            &            &         38 \\
A 68 & 060.0-04.3 &         37 &         30 &            &        yes &         HE &            &            &            \\
A 71 & 084.9+04.4 &         22 &         18 &            &        yes &         HE &       2.37 &       5.09 &         29 \\
A 72 & 059.7-18.7 &            &         54 &            &         no &          M &            &            &        110 \\
A 73 & 095.2+07.8 &         33 &         26 &            &        yes &         HE &            &            &            \\
A 75 & 101.8+08.7 &            &         39 &            &         no &          M &       3.28 &        4.9 &         81 \\
A 80$^1$ & 102.8-05.0 &     23 &         18 &            &        yes &         HE &       1.79 &       5.09 &         25 \\
A 82 & 114.0-04.6 &         30 &         25 &            &        yes &         HE &            &            &         15 \\
A 83 & 113.6-06.9 &        32 &         25 &            &        yes &         HE &             &            &            \\
A 84$^1$ & 112.9-10.2 &     28 &         24 &            &        yes &         HE &       1.94 &          5 &         17 \\
DeHt 2 & 027.6+16.9 &          &         48 &            &         no &         HE &            &            &            \\
DS1 &  283.9+9.7 &             &         19 &         18 &        yes &         HE &       3.48 &       4.95 &            \\
EGB 1 & 124.01+10.7 &        9 &         14 &            &        yes &         HE &       2.37 &       5.17 &            \\
HDW 2 & 138.1+04.1 &        17 &         15 &            &        yes &         HE &            &            &            \\
HDW 3 & 149.4-09.2 &       18 : &            &            &        yes &         HE &           &            &            \\
HDW 6$^1$ & 192.5+07.2 &     21 &         18 &            &        yes &         HE &           &            &            \\
HDW 7 & 211.4+18.4 &            &         21 : &            &         no &         HE &         &            &            \\
HFG 1$^2$ & 136.3+05.5  &      &            &       18 : &     [OIII] &         HE &       3.33 &          5 &            \\
IC 1295 & 025.4-04.7 &      34 &         31 &            &        yes &         HE &        2.4 &       4.99 &         50 \\
IC 972 & 326.6+42.2 &       25 &         20 &            &        yes &         HE &            &            &         20 \\
Jn 1 & 104.2-29.6 &            &         40 : &            &        yes &       HE &       2.59 &       5.18 &         50 \\
JnEr1& 164.8+31.1 &         35 &         31 &            &        yes &         HE &       2.02 &       5.06 &            \\
K 1-1 & 252.6+04.4 &        20 &         19 &            &        yes &         HE &            &            &         33 \\
K 1-11 & 215.6+11.1 &       40 &         31 &            &        yes &         HE &            &            &            \\
K 1-12$^2$ & 236.7+03.5 &   23 &      22  * &            &        yes &          M &            &            &            \\
K 1-13 & 224.3+15.3 &       47 &         42 &            &        yes &         HE &            &            &            \\
K 1-14 & 045.6+24.3 &          &            &         58 &         no &          M &            &            &        155 \\
K 1-20 & 110.6-12.9 &       37 &         33 &            &        yes &          M &            &            &         10 \\
K 1-22 & 283.6+25.3 &       40 &         33 &            &        yes &         HE &        2.3 &       5.06 &         10 \\
K 1-28 & 270.1+24.8  &         &         30 &            &         no &          M &            &            &            \\
K 1-3 & 346.9+12.4   &      22 &         20 : &            &       yes&         HE &            &            &          25\\
K 2-1 & 173.7-05.8  &          &         34 &            &         no &         HE &       2.63 &       5.05 &        115 \\
K 2-16 & 352.9+11.4 &       30 &         29 &            &        yes &         HE &            &            &            \\
K 2-5 & 014.9+06.4 &        37 &         27 &            &        yes &          M &            &            &            \\
Kn 34 & 149.1+08.8 &        38 &         35 &            &        yes &         HE &            &            &            \\
Kn 36$^2$ & PASA2010-1 &    30 &         26 &            &        yes &         HE &            &            &            \\
Kn 37 & 165.5-05.2 &           &         26 &            &         no &         HE &            &            &            \\
Kn 39$^2$ & PASA2010-2 &       &     31 : * &            &        yes &         HE &            &            &            \\
Kn 40 & 198.9-06.1 &           &         32 &            &         no &         HE &            &            &            \\
Lo 11 & 340.8+12.3 &        48 &         47 &            &        yes &         HE &            &            &            \\
Lo 12 & 340.8+10.8 &        34 &         33 &            &        yes &         HE &            &            &            \\
Lo 13 & 345.5+15.1 &           &         20 &            &         no &         HE &            &            &            \\
Lo 8 & 310.3+24.7 &            &       50 : &            &         no &         HE &       3.89 &       4.95 &            \\
LoTr 1 & 228.2-22.1 &          &            &         22 &        yes &         HE &            &            &            \\
LoTr 5 & 339.9+88.4 &          &            &         31 &     [OIII] &         HE &       2.01 &          5 &            \\
NGC 1360$^2$ & 220.3-53.9 &    &         28 &            &         no &          M &        3.3 &       5.04 &        120 \\
NGC 1501 & 144.1+06.1 &     41 &         39 &            &        yes &          M &       3.66 &       5.13 &         70 \\
NGC 246 & 118.7-74.7  &        &         40 &         37 &         no &          M &       3.63 &       5.15 &        100 \\
NGC 3587 & 148.4+57.0 &     40 &         34 &            &        yes &         HE &       1.95 &       5.02 &         14 \\
NGC 7094 & 066.7-28.2 &        &         38 &            &         no &          M &       3.61 &       5.04 &        120 \\
NGC 7139 & 104.1+07.9 &     38 &         33 &            &        yes &          M &            &            &         14 \\
NGC 7293 & 036.1-57.1  &    21 &            &            &        yes &         HE &       1.95 &       5.04 &         10 \\
Pa 5 & 076.3+14.1 &            &         42 &         46 &         no &         HE &            &            &            \\
Pa 9 & 189.1-07.7 &           38 : &         37 &            &         no &         HE &            &            &            \\
PHR J0843-2514 & 248.5+10.5 &     &         31 &         32 &     no &         HE &         &            &            \\
PHR J1710-3732 & 348.7+01.3 &  31 &         29 &            &        yes &         HE &         &            &            \\
PHR J1715-2905 & 356.2+05.3 &     &            &         51 &     [OIII] &         HE &         &            &            \\
PHR J1734-2000 & 006.2+06.9 &     &         40 &            &         no &         HE &         &            &            \\
PHR J1754-3533$^2$ & 355.2-05.0 &  &       30 * &            &         no &         HE &        &            &            \\
Pu 1 & 181.5+00.9 &         21 &       22 * &            &        yes &         HE &            &            &         20 \\
Pu 2 & 173.5+03.2 &            &         23 &         22 &         no &         HE &            &            &            \\
PuWe 1$^2$ & 158.9+17.8 &   30 &         34 &            &        yes &         HE &        1.7 &       5.04 &         37 \\
Ri 1 & 204.8+02.7 &         26 &         26 &            &        yes &         HE &            &            &            \\
Sa 2-21 & 238.9+07.3 &      38 &         29 &            &        yes &          M &            &            &            \\
SaWe 1 & 233.0-10.1 &          &         32 &            &         no &         HE &            &            &            \\
Sb 32 & 349.7-09.1 &        38 &         33 &            &        yes &         HE &            &            &            \\
We 1-1 & 121.6+03.5 &       31 &         22 &            &        yes &         HE &            &            &         15 \\
We 1-10 & 086.1+05.4 &     27 : &            &            &        yes &         HE &           &       4.76 &            \\
We 1-2 & 160.5-00.5 &       26 &         18 &            &        yes &         HE &            &            &            \\
We 1-3$^2$& 163.1-00.8 &    28 &            &            &        yes &         HE &            &            &            \\
We 1-4 & 201.9-04.6 &       18 &         21 &            &        yes &         HE &            &            &            \\
We 1-5 & 216.3-04.4 &          &         34 &            &         no &          M &            &            &         95 \\
WeSb 1 & 124.3-07.7 &       27 &         25 &            &        yes &         HE &            &            &            \\
Wray16-22 & 255.7+03.3 &    39 &         34 &            &        yes &         HE &            &            &            \\
YM16 & 038.7+01.9 &         25 &         26 &            &        yes &         HE &            &            &            \\

\enddata
\tablenotetext{a}{The HE and M classifications correspond to Highly Evolved and Mature objects respectively.}
\tablenotetext{b}{Data taken from \cite{Frew08}.}
\tablenotetext{c}{The I(HeII$\lambda$4686) data were taken from \cite{Tylenda94}.}
\tablenotetext{1}{Slit position is not at center of the PN. Lower limit for V$_{flow}$.}
\tablenotetext{2}{Upper limit for V$_{flow}$.}
\tablenotetext{*}{V$_{flow}$ measurement obtained from the FWHM of a Gaussian fit.}
\tablenotetext{:}{Very faint PN. Velocity measurements must be treated carefully.}

\end{deluxetable}

\clearpage

\end{document}